\documentclass[prl,twocolumn,showpacs,amssymb]{revtex4}

\usepackage{graphicx}

\input psfig
\begin{document}

%\usepackage{amsmath}

%\twocolumn        

%\twocolumn[\hsize\textwidth\columnwidth\hsize\csname
%@twocolumnfalse\endcsname

\newcommand*{\be}{\begin{equation}} 
\newcommand*{\ee}{\end{equation}}

%\def\Zsun{\thinspace\hbox{$\hbox{Z}_{\odot}$}}
%\def\msun{\thinspace\hbox{$\hbox{M}_{\odot}$}}
%\def\rsun{\thinspace\hbox{$\hbox{R}_{\odot}$}}
%\def\lsun{\thinspace\hbox{$\hbox{L}_{\odot}$}} 
%\def\.{\'{\i}}

%  The most r

\title{Hubble diagram of gamma-ray bursts: Robust evidence for a Chaplygin gas 
expansion-driven universe with phase transition at $z \simeq 3$ }

\author{Herman J. Mosquera Cuesta$^{1,2}$, Habib Dumet M.$^{1}$, Rodrigo Turcati$^{1}$,  
Carlos A. Bonilla Quintero$^{1}$, Cristina Furlanetto$^{1}$, and Jefferson Morais$^{1}$}

\address{$^1$Instituto de Cosmologia, Relatividade e Astrof\'{\i}sica 
(ICRA-BR), Centro  Brasileiro de Pesquisas F\'{\i}sicas \\ Rua Dr. Xavier 
Sigaud 150, CEP 22290-180  Urca, Rio de Janeiro, RJ, Brazil \\
$^2$\mbox{Abdus Salam International Centre for Theoretical Physics, Strada 
Costiera 11, Miramare 34014, Trieste, Italy } }

\date{\today}

\begin{abstract} 
{
The Hubble diagram (HD) of Gamma-Ray Bursts (GRBs) having properly estimated redshifts is 
compared with the predicted one for the Chaplygin gas (CG), a dark energy  candidate. The 
CG cosmology and that of Friedmann and $\Lambda$-CDM models are studied and confronted to 
the GRBs observations. The model-to-sample $\chi^2$ statistical analysis indicates the CG
model as the best fit. The present GRBs HD plot exhibits a marked trend: as one goes back 
in time, it gets much closer to the predict HD for a Friedmann universe. This clear trend 
conclusively demonstrates that a transition from decelerate to accelerate expansion did 
take place. However, contrarily to claims based on supernovae type Ia, the transition 
redshift lies somewhere between $\sim 2.5 < z \simeq 3.5$ rather than at $z \sim 0.5-1$. 
All of these striking features of the GRBs HD constitutes the most robust demonstration 
that the Chaplygin gas can in fact be the universe's driving dark energy field.
}

%%  as the actual stand-point for identifying the transition to CG dominated universe.

%%and since the GRBs already span a redshift range corresponding to factually ver large cosmic scales, 
\end{abstract}


\email[E-mail me at: ]{hermanjc@cbpf.br}

\maketitle

%\tableofcontents

{\sl Introduction.---} 
Observations of supernovae type Ia (SNIa) have led to the current view that our 
universe underwent a late-time transition to accelerate expansion at a redshift 
$z\sim 1$. The driver of such unexpected dynamics is an exotic component of the 
universe's content dubbed dark energy (DE), a smoothly sparsed energy field with 
no familiar counterpart among the currently known forms of matter-energy. Lots of 
theories have been conceived to explain this striking phenomenon in contemporary 
cosmology. One of these candidates for DE is the so-called Chaplygin gas (CG), a
strange fluid described by a quite unsual equation of state (EoS)
\be
p = - \frac{A}{\rho} 
\label{CG} ,
\ee 
where $p$ represents the pressure, $\rho$ the fluid density and $A$ is a constant. 
It came after the russian aerodynamicist Chaplygin who in 1904 brought it in to explain 
the lifting force on a plane wing in some aerodymamic phenomena \cite{chaplygin1904}. By 
2001, Kamenschik, Moschella and Pasquier \cite{moschella01} recognised its relevance to 
cosmological studies, in particular, with respect to the claimed cosmic acceleration. 
They showed that the CG model exhibits excelent agreement with observations. Besides, 
the model predicts a larger value for the effective cosmological constant. The same 
researchers noticed then that the model can be generalized in the form: $p = - \frac{A} 
{\rho^\alpha}$, and consider the case with the density power-law exponent $\alpha = 1/3$ 
\cite{moschella01}. It was then realized that the CG EoS has a clearly stated connection 
with string and brane theories\cite{moschella00,bertolami06,jackiw00}.\footnote{ From now 
on we will refer to the generalized Chaplygin gas (GCG) model with the $\alpha$-parameter 
since it will be the base of a detailed study that will be presented in the accompanying 
paper to this {\sl Letter}\cite{nos2006a}. Nonetheless, in the discussion ahead we will 
restrict ourselves to the case wherein $\alpha = 1$.}

\begin{figure}[tbh]
\centering{
%\resizebox{8.2cm}{!}{
%[angle=-90]
%
\includegraphics[height=2.75in,width=3.0in]{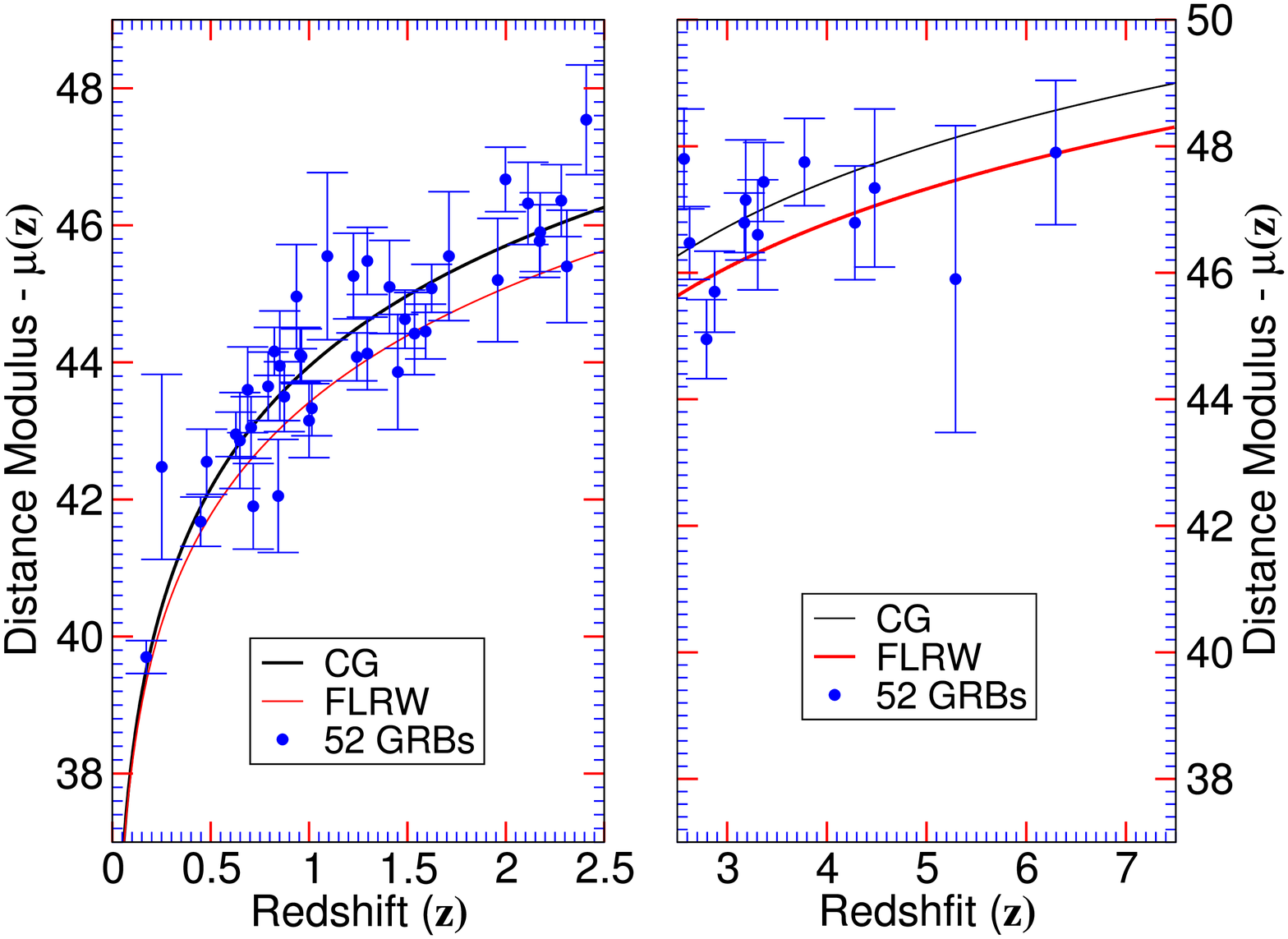} }
\caption{(color on-line) HD of GRBs and the CG model of DE. The plot demonstrates that a
phase transition did take place at a redshift $ 2.5 \lesssim z \simeq 3.5$, or even earlier 
(right panel) } 
\label{figure1}
\end{figure}

Here we build the Hubble diagram (HD) of gamma-ray bursts (GRBs)\footnote{Gamma-ray 
bursts (GRBs) are the biggest explosions in the universe. The 1997 february 28 
Beppo-SAX discovery of the X-ray afterglow of a gamma-ray burst (GRB) allowed the 
first precise determination of the redshift of a GRB. This major breakthrough came to 
confirm the long-standing suspicion waving in the high energy astrophysics community 
that GRBs arrive from, and their sources lie at, cosmological whereabouts. Since then, 
the possibility of using them as actual cosmological probes has estimulated the search 
for self-consistent methods of bringing GRBs into the realm of cosmology. A handful of 
attempts have been on trial. Since the introduction of the {\sl Amati relation} 
\cite{Amati2002}, other largely promising techniques have appeared in an attempt 
to turn GRBs into reliable cosmological probes. These include the {\sl Ghirlanda relation} 
\cite{Ghirlanda2004}, the {\sl Liang-Zhang relation} \cite{Liang-Zhang2005}, and the
recently discovered {\sl Firmani et al. relation} \cite{Firmani-etal.-2006a},  all of 
which taking into account the most relevant physical properties of GRBs as the peak energy, 
jet openning angle, and both time lag and variability. Such discoveries hints at the 
long-sought {\sl Holy Grail} of creating a cosmic ruler from  GRBs observables to be 
achievable. Presently, after the first series of controversial statements on the viability 
of granting to GRBs the status of standard cosmic rulers 
\cite{Schaefer2003,Bloom-Frail-Kulkarni2003,Friedman-Bloom2005}, a definite consensus 
appears to be arising, and the hope to have a {\sl Holy Grail} to do cosmology upon GRBs is 
renascenting 
\cite{Ghirlanda2004,Friedman-Bloom2005,Mortsell-Sollerman-2005,Schaefer2006}. Besides, we 
bring to the reader's attention a historical fact that has a clear correspondance with the 
present situation regarding the GRBs cosmology. Despite having still large error bars, and 
in several cases a not so clear estimate of the redshift of some events, the present state 
of the cosmology based on GRBs emulates the days during which Hubble discovered the 
expansion of the universe by using inhomogeneous and badly calibrated data from the nebulae 
he, Humason and others had observed \cite{hubble1929}. Nobody nowadays thinks of that his 
analysis as null and void, in spite of his methods having been not so standardized. Keeping
these arguments in mind one can be confident of the worthiness for cosmological studies of 
the analysis being presented here.}, which already reach redshift $z \sim 7$, and SNIa, 
and confront it with predictions of the cosmological model in which our 
universe is filled-in with this sort of DE, the so-called Chaplygin gas. In comparing 
the CG predicted HD with both GRBs and SNIa data is verified that the best fit to the 
observational data clearly correspond to this exotic fluid, when compared to the HD 
for the Friedmann-Lema\^itre-Robertson-Walker (FLRW, the universe contents only matter) 
and the Lambda Cold Dark Matter ($\Lambda$-CDM, Friedmann cosmology with $\Lambda$) 
scenarios. Three major achievements are attained in this investigation:
a) for the first time is presented the HD of GRBs in confrontation to the CG theoretical 
prediction, and that of the FLRW and $\Lambda$-CDM models (see FIG. 1). b) it is shown 
that the best 
fit ($\chi^2$-statistics) to the GRBs data is provided by the Chaplygin gas scenario. c) 
the resulting HD clearly exhibits a transition from a Friedmann-dominated to a late-time 
accelerating universe, with the transition taking place at a redhsift around $ 2.5 \lesssim 
z \simeq 3.5$ (see FIG. 1), and driven by the CG. Besides, the similar analysis for SNIa 
allows one to verify that the HD for SNIa clearly violates the CG, FLRW and $\Lambda$-CDM 
expansion law. This disagreement with the Chaplygin gas HD, once again suggests that perhaps 
there is something wrong with the SNIa observations and that its astrophysics deserves to be 
revisited.

\begin{figure*}[th]
\centering{
%\resizebox{8.2cm}{!}{
%[angle=-90]
%
\includegraphics[height=2.0in,width=2.0in]{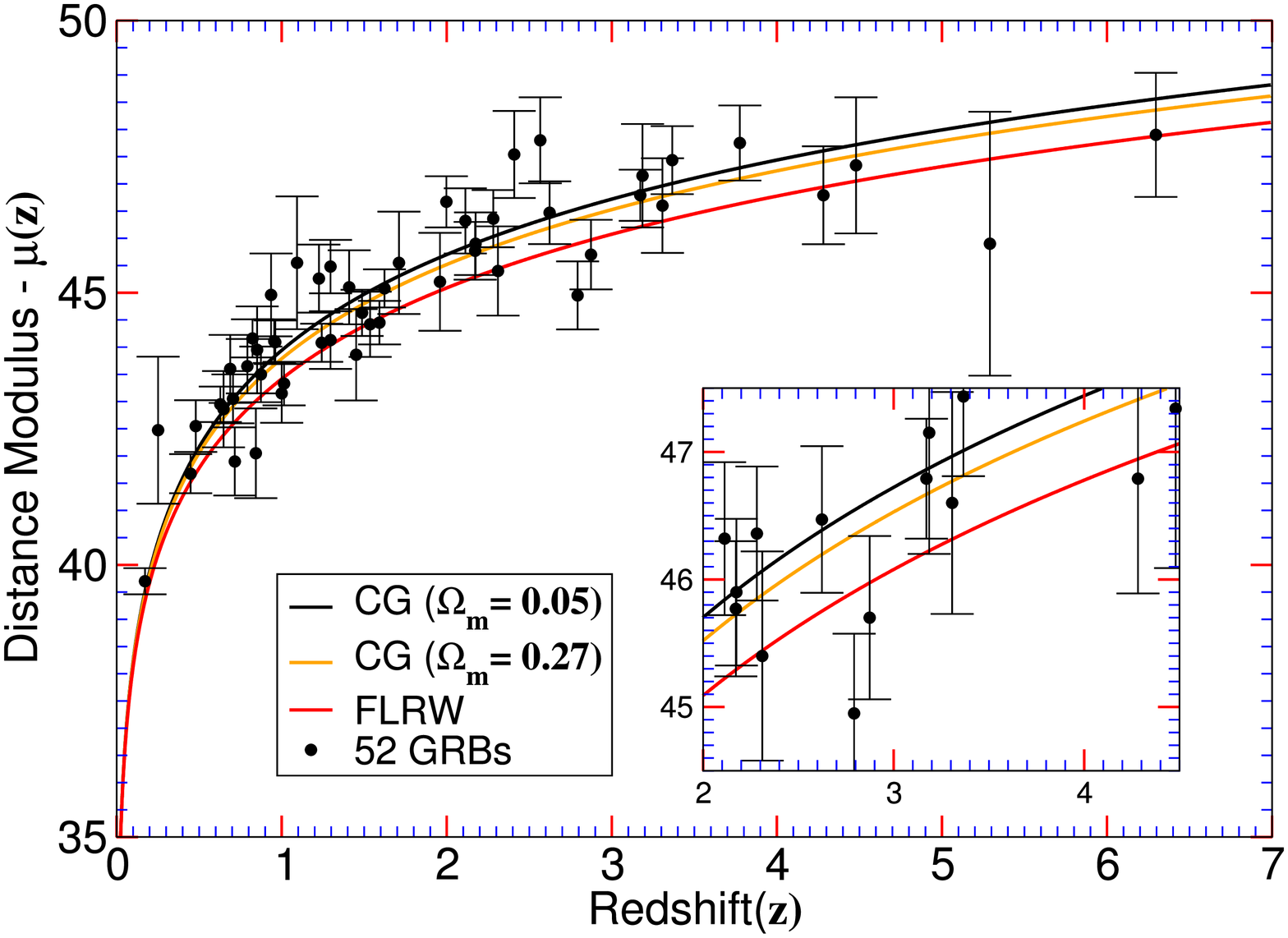} \hskip 0.40truecm 
\includegraphics[height=2.0in,width=2.0in]{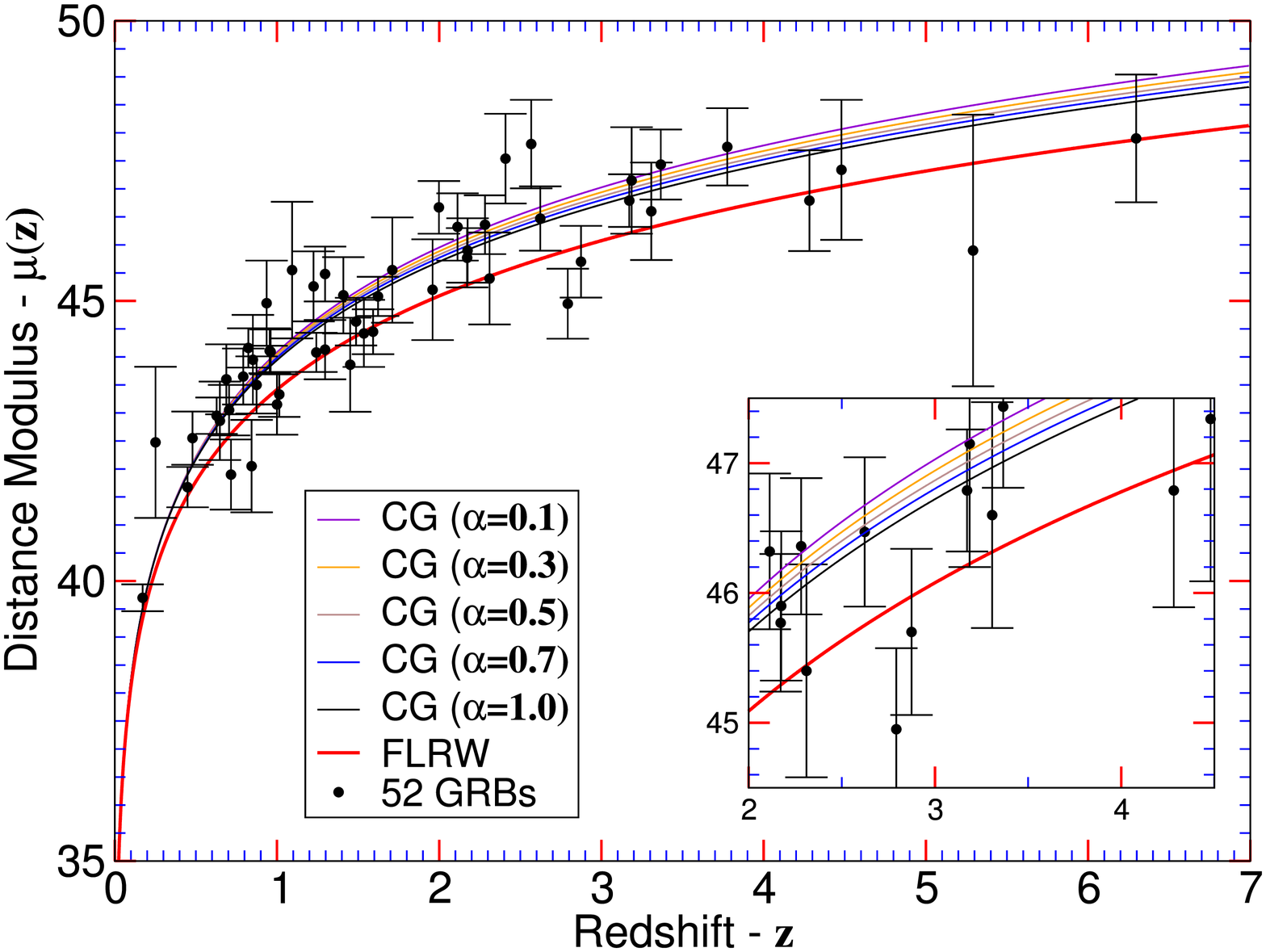} \hskip 0.40 truecm 
\includegraphics[height=2.0in,width=2.0in]{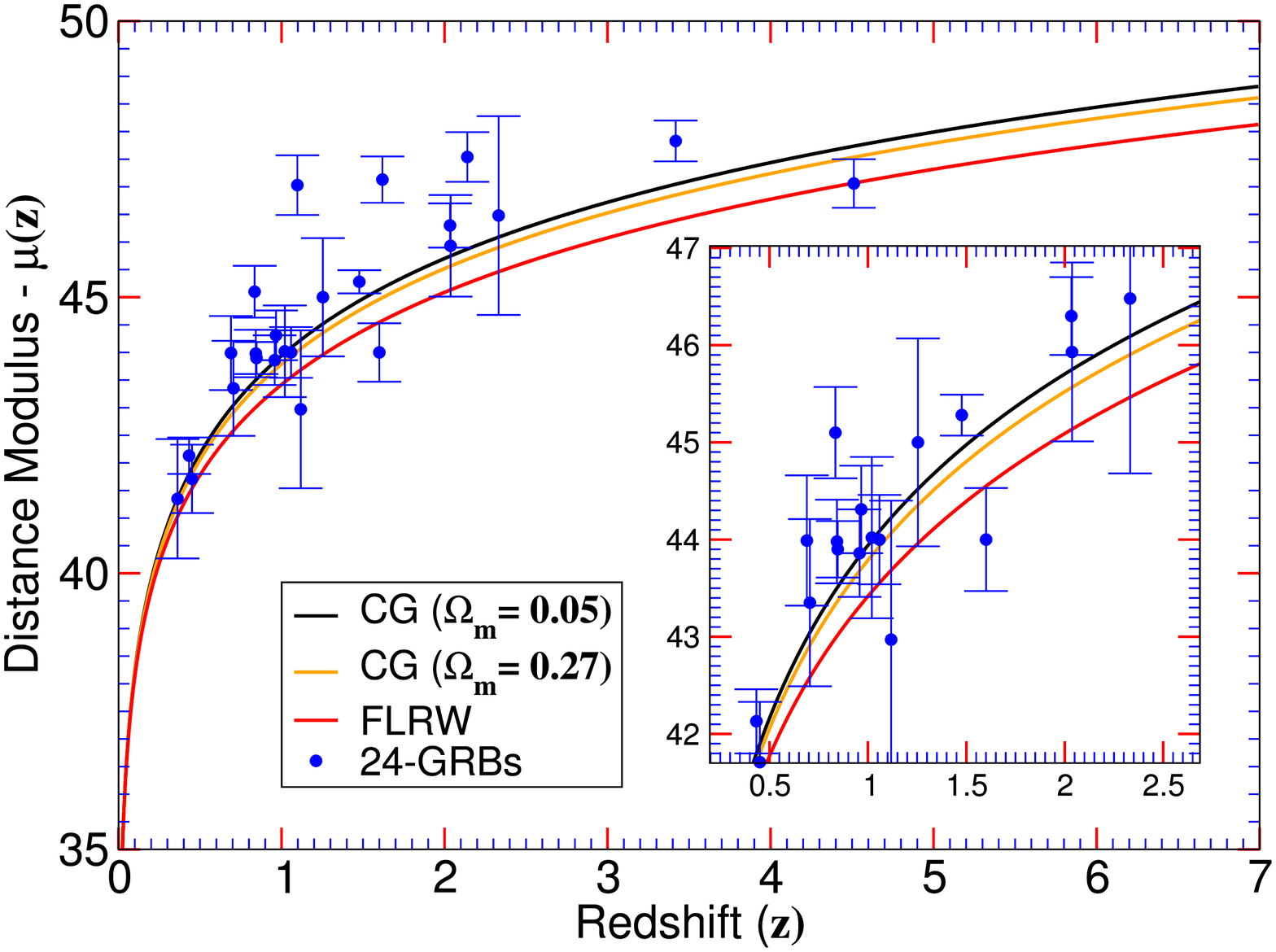}\vskip 0.50 truecm
\includegraphics[height=2.0in,width=2.0in]{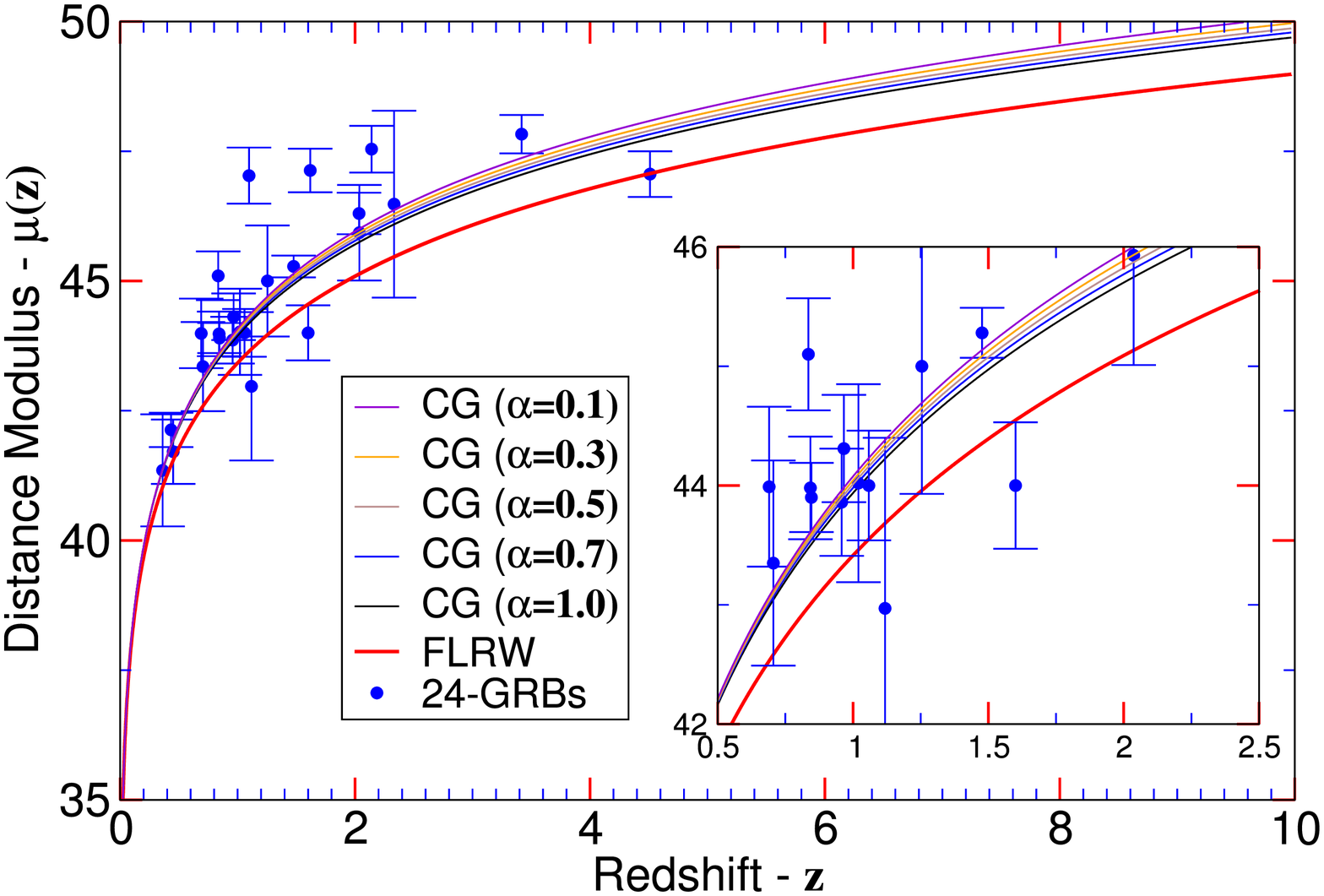}\hskip 0.40 truecm
\includegraphics[height=2.0in,width=2.0in]{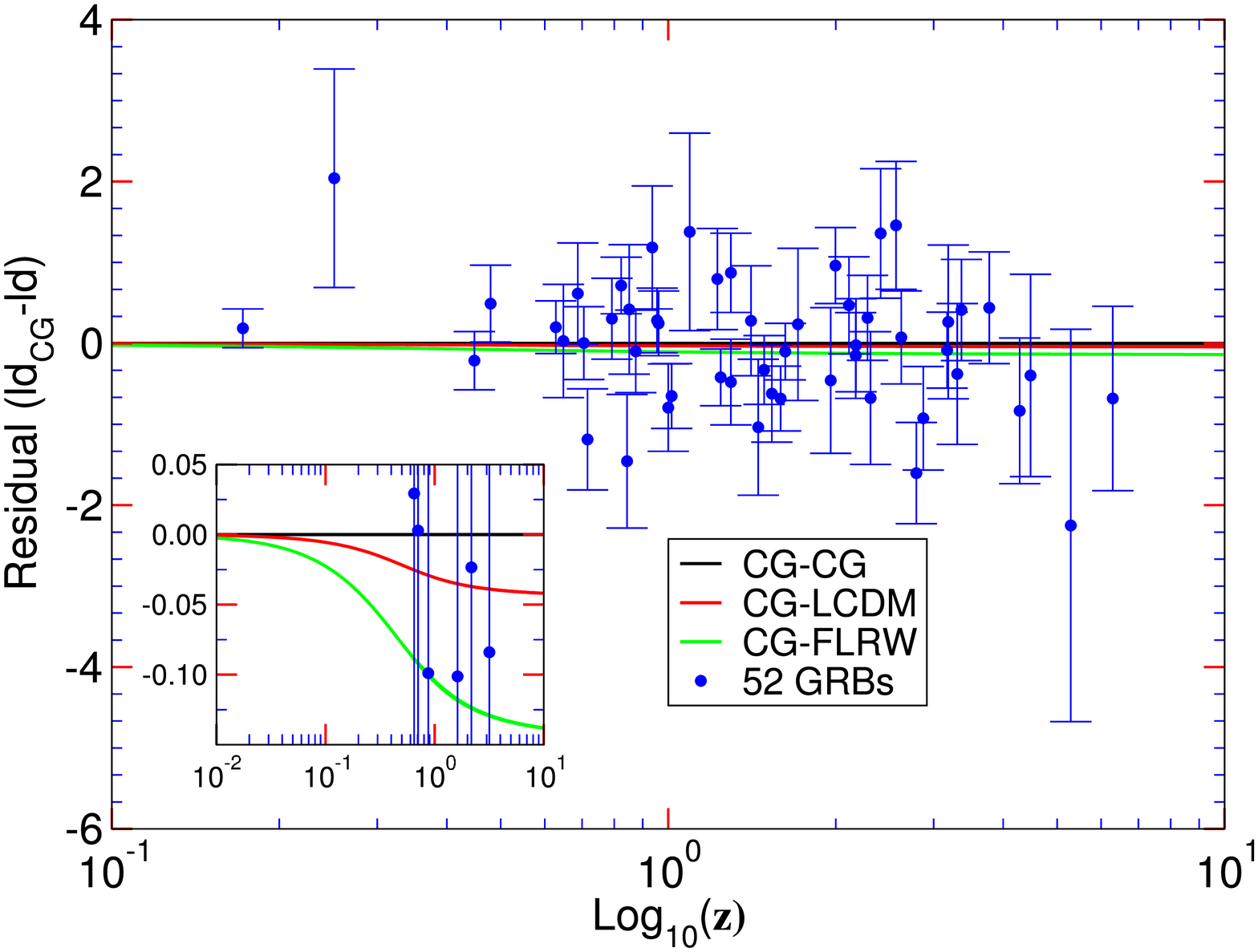}\hskip 0.40 truecm
\includegraphics[height=2.0in,width=2.0in]{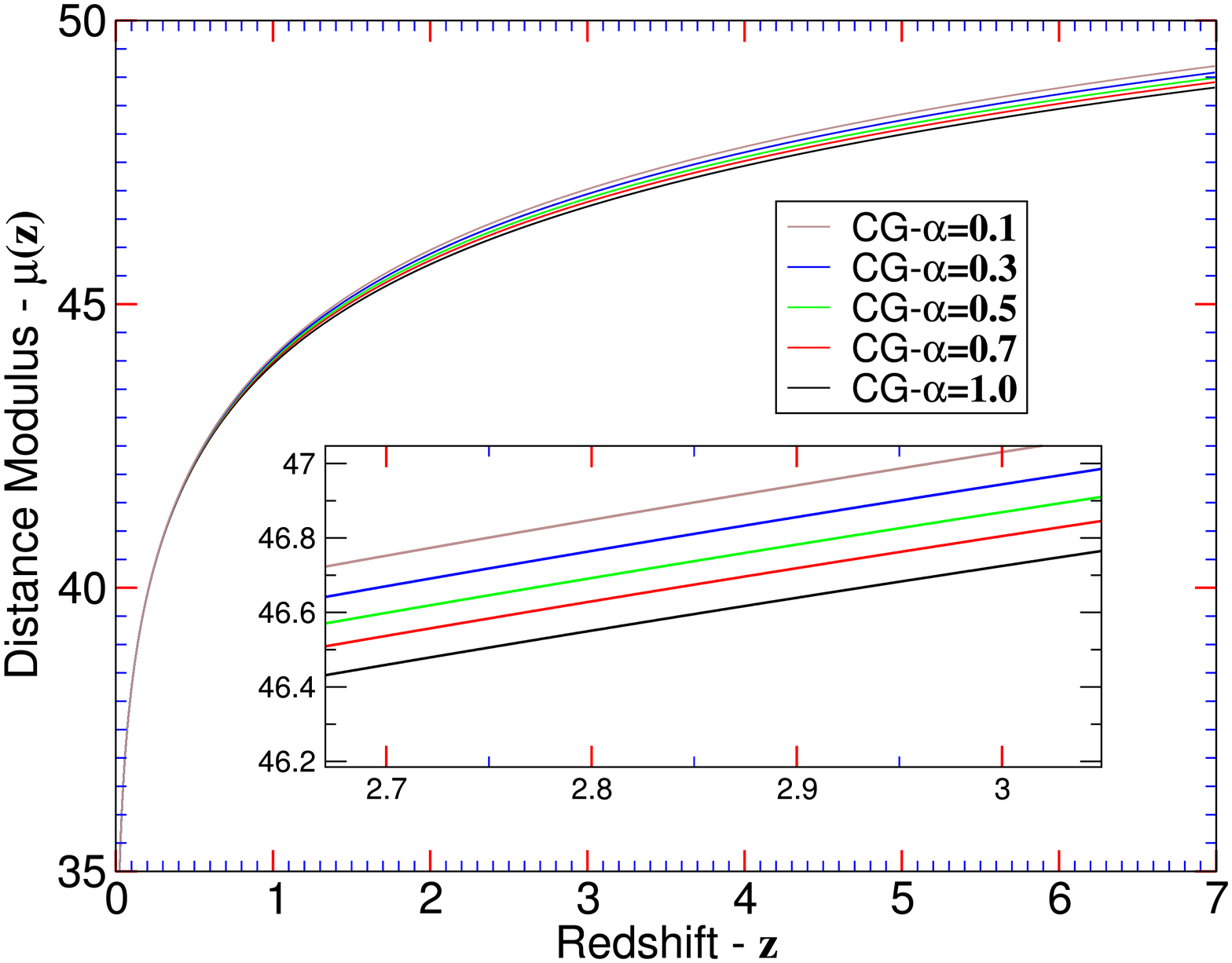} }
\caption{(color on-line) HD of GRBs and the CG model of DE for the 52 (first pair of plots)
and 24 GRBs (second pair) samples, including the case of a cosmological constant with 
$\Omega_\Lambda = 0.73$. Also the residual HD of the CG vs. Friedmann models confronting 
the 52 GRBs sample, and the HD small variation for several values of the CG 
$\alpha$-parameter are presented. } 
\label{figure3}
\end{figure*}

{\sl Chaplygin gas cosmology.---} 
The enthusiasm about the Chaplygin gas dynamics stems from the possibility of unifying 
dark matter and dark energy in a cosmic fluid that could be represented by a scalar 
field. As one can notice from the evolution law for its energy density  
\cite{moschella01,bertolami04}
\begin{equation}
\quad \rho_c = \biggr(A + \frac{B}{a^{3(1 + \alpha)}} \biggl)^{1/(1 + \alpha)} \quad ,
\label{CG-density}
\end{equation}
where $a$ is the scale factor of the universe and $B$ is an integration constant,
the CG corresponds to the $\Lambda$-CDM scenario for the parameter $\alpha = 0$, 
and $A_s \equiv A/ \rho^{1 + \alpha}_{c0} = 1$. A direct analysis 
of Eq.(\ref{CG-density}) shows that the present acceleration phase, which should 
have taken place at a redshift 
\be
z_{t} = \left([3\omega(z_t) + 1] \frac{[\Omega_m - 1]}{\Omega_m} \right)^{\frac{- 1}
{\hat{\omega}(z_t)}} - 1  \; , 
\label{transition-redshift}
\ee
where $\hat{\omega}(z_a) = \frac{1}{ \ln(1 + z)} \int^z_0 \frac{\omega(z^{'})}{1 + 
z^{'}} dz^{'}$, leads to an asymptotic ($a \rightarrow \infty$) stage where the EoS 
is dominated by a cosmological constant ($ 8 \pi A^{1/1+\alpha}$), whereas at earlier 
epochs the energy density is dominated by a non-relativistic matter. Notice also that 
the value $a_{t}$\ of the scale factor that signalizes the start of the late-time 
acceleration phase is given by the roots of the equation \cite{fabris06a}
\begin{equation}
\ddot{a}=a(\dot{H}+H^{2})\quad ,
\end{equation}
and is related to the redshift value $z_{t}$ such that $\frac{a_{t}}{a_{0}} = 
\frac{1}{1+z_{t}}$.

The highlighted dual behavior is the basis of the unification scheme provided by 
the CG model. This unification becomes possible if one takes benefit of a complex 
scalar field $\Phi$ of mass $m_\phi$ that admits an inhomogeneous generalization, 
as demanded by models of structure formation in the universe, and is described by 
a Lagrangian density \cite{moschella01,bertolami04}
\be
{\cal L} = g_{\mu \nu} \bar{\Phi}^\mu \Phi^\nu - V(|\Phi|)^2
\label{lagrangian}
\ee
where $\Phi = (\phi/\sqrt{2}~ m ) \exp(-i m_\phi \theta)$. The cosmic dynamics of this 
field was developed in Ref.\cite{moschella01,bertolami04}.

%%%%%%%%%%%%%%%%%%%%%%%%%%%%%%%%%%%%%%%%%%%%%%%%%%%%%%%%%%%%%%%%%%%%%%%%%%%%%%%%%%%%%%

%%%%%%%%%%%%%%%%%%%%%%%%%%%%%%%%%%%%%%%%%%%%%%%%%%%%%%%%%%%%%%%%%%%%%%%%%%%%%%%%%%%%%%%%%

As pointed out above, the GCG Model can be idealized by a perfect fluid with an EoS 
given by
\begin{equation}
p = - \frac{A}{\rho^\alpha} \quad ,
\label{GCG}
\end{equation}
where $A$ and $\alpha$ are constants. When $\alpha = 1$ we
re-obtain the EoS for the CG scenario (Eq.\ref{CG}). In principle, 
the parameter $\alpha$ is restricted in such a way that $0 \leq 
\alpha \leq 1$. However, possible values for $\alpha \neq (0,1]$ 
are considered in the accompanying paper \cite{nos2006a}, where we 
also analyze the effects of imposing the energy conditions to the 
cosmic dynamics of the CG as to be compared with GRBs and SNIa 
observations.

The universe content can be envisioned as having a pressureless matter, needed to
account for the presence of baryons in it, and also dark matter (also pressureless) 
and dark energy making-up the CG. Hence, the dynamics of the Universe is worked out 
through the Friedmann's equation and the evolution equations for non-interacting 
baryonic matter and Chaplygin gas\cite{fabris06a}
\begin{eqnarray}
&&\biggr(\frac{\dot a}{a}\biggl)^2 + \frac{k}{a^2} = \frac{8\pi G}{3}\biggr(
\rho_m + \rho_c\biggl) \quad , \label{be1} \\
&&\dot\rho_m + 3 \frac{\dot a}{a}\rho_m = 0 \quad , \label{be2} \\
&&\dot\rho_c + 3\frac{\dot a}{a}\biggr(\rho_c - \frac{A}{\rho_c^\alpha}\biggl) 
= 0 \quad \label{be3} ,
\end{eqnarray}
where $\rho_m$ and $\rho_c$ stand for the pressureless matter and Chaplygin
gas component, respectively. As usual, $k = 0, 1, - 1$ indicates a flat,
closed and open spatial section.

The conservation law for each of these fluids 
(\ref{be2},\ref{be3}) reads: $\rho_m = \frac{\rho_{m0}}{a^3}\; $, and 
$\rho_c = (A + \frac{B}{a^{3(1 + \alpha)}})^{1/(1 + \alpha)}\;
\label{evolution-rho-matter-chaplygin}$, respectively.
The value of the scale factor today is taken equal to unity, $a_0 = 1$.
Hence, $\rho_{m0}$ and $\rho_{c0} = (A + B)^{1/(1 + \alpha)}$
are the pressureless matter and GCG densities today. Eliminating from the
last relation the parameter $B$, the GCG density at any time can be
re-expressed as
\begin{equation}
\rho_c = \rho_{c0}\biggr(\bar A + \frac{1 - \bar A}{a^{3(1 + \alpha)}} %
\biggl)^{1/(1 + \alpha)} \quad ,
\end{equation}
where $\bar A = A/\rho_{c0}$. This parameter $\bar A$ is connected with the
sound velocity for the Chaplygin gas today by the relation $\frac{\partial p}
{\partial \rho} = v_s^2 = \alpha\bar A$.

Our main purpose here is the theoretical distance modulus vs. redshift relation, 
i.e., the HD of the CG model, to compare it with GRBs observations. For this, we 
need the luminosity distance \cite{weinbergb,coles}
\begin{equation}
d_L = \frac{a_0^2}{a}~r_1 \quad ,
\end{equation}
with $r_1$ the co-moving coordinate of the source. As light propagates on a null 
geodesic, i.e.,
\begin{equation}
ds^2 = c^2 dt^2 - \frac{a^2dr^2}{1 - kr^2} = 0 \quad ,
\end{equation}
the Friedmann's equation (\ref{be1}) allows to re-cast the luminosity distance as
\begin{equation}
d_L = (1 + z)S[f(z)] \quad ,
\label{luminosity-distance}
\end{equation}
where $S(x) = x \quad (k = 0)$, $S(x) = \sin x \quad (k = 1)$, $S(x) = \sinh x 
\quad (k = - 1)$ , and the function $f(z)$ being given by 
\begin{widetext}
\begin{equation}
f(z) = \frac{c}{H_0}\int_0^z \frac{d\,z^{\prime
}}{\{\Omega_{m0}(z^{\prime}+ 1)^3 + \Omega_{c0}[\bar A +
(z^{\prime}+ 1)^{3(1+\alpha)}(1 - \bar A)]^{1/(1+\alpha)} -
\Omega_{k0}(z^{\prime}+ 1)^2\}^{1/2}} \quad ,
\label{luminosity-distance-function}
\end{equation}
\end{widetext}
with the definitions
\begin{equation}
\Omega_{m0} = \frac{8\pi G}{3}\frac{\rho_{m0}}{H_0^2} \; ,
\Omega_{c0} = \frac{8\pi G}{3}\frac{\rho_{c0}}{H_0^2} \; ,
\Omega_{k0} = - \frac{k}{H_0^2} \; ,
\end{equation}
such that the condition $\Omega_{m0} + \Omega_{c0} + \Omega_{k0} = 1$ holds. The final 
equations did use of the redshift vs. scale factor relation: $1 + z = \frac{(a_0=) 1}{a}$.

%The age of the Universe and the value of the decelerated parameter $q_0 = -
%\frac{a\ddot a}{\dot a^2}$ are given by
%\begin{widetext}
%\begin{eqnarray}
%\frac{t_{0}}{T}&=&\int_{0}^{z}\frac{d\,z^{\prime }}{(1+z^{\prime })\{\Omega
%_{m0}(z^{\prime }+1)^{3}+\Omega _{c0}[\bar{A}+(z^{\prime }+1)^{3(1+\alpha
%)}(1-\bar{A})]^{1/(1+\alpha )}-\Omega _{k0}(z^{\prime }+1)^{2}\}^{1/2}}\quad
%, \\
%q_{0} &=& \frac{\Omega _{m0}+\Omega _{c0}(1-3\bar{A})}{2} ,
%\end{eqnarray}
%\end{widetext}
%where $T=(100/H_{0})\times 10^{10}$, so that $t_{0}$ has units of
%years.

\begin{figure*}[tbh]
\centering{
%\resizebox{8.2cm}{!}{
%[angle=-90]
\includegraphics[height=2.0in,width=2.0in]{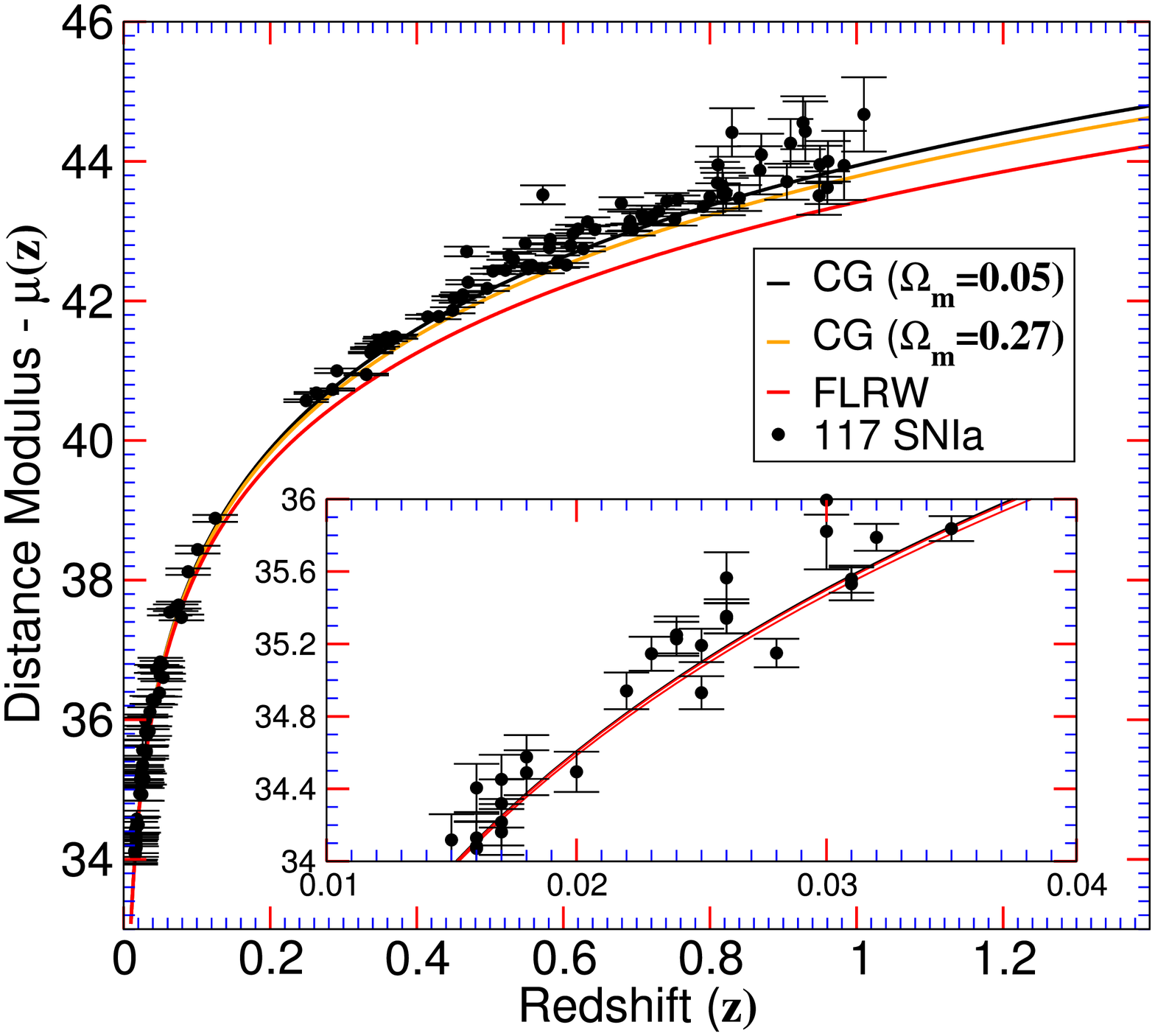}\hskip 0.40truecm 
\includegraphics[height=2.0in,width=2.0in]{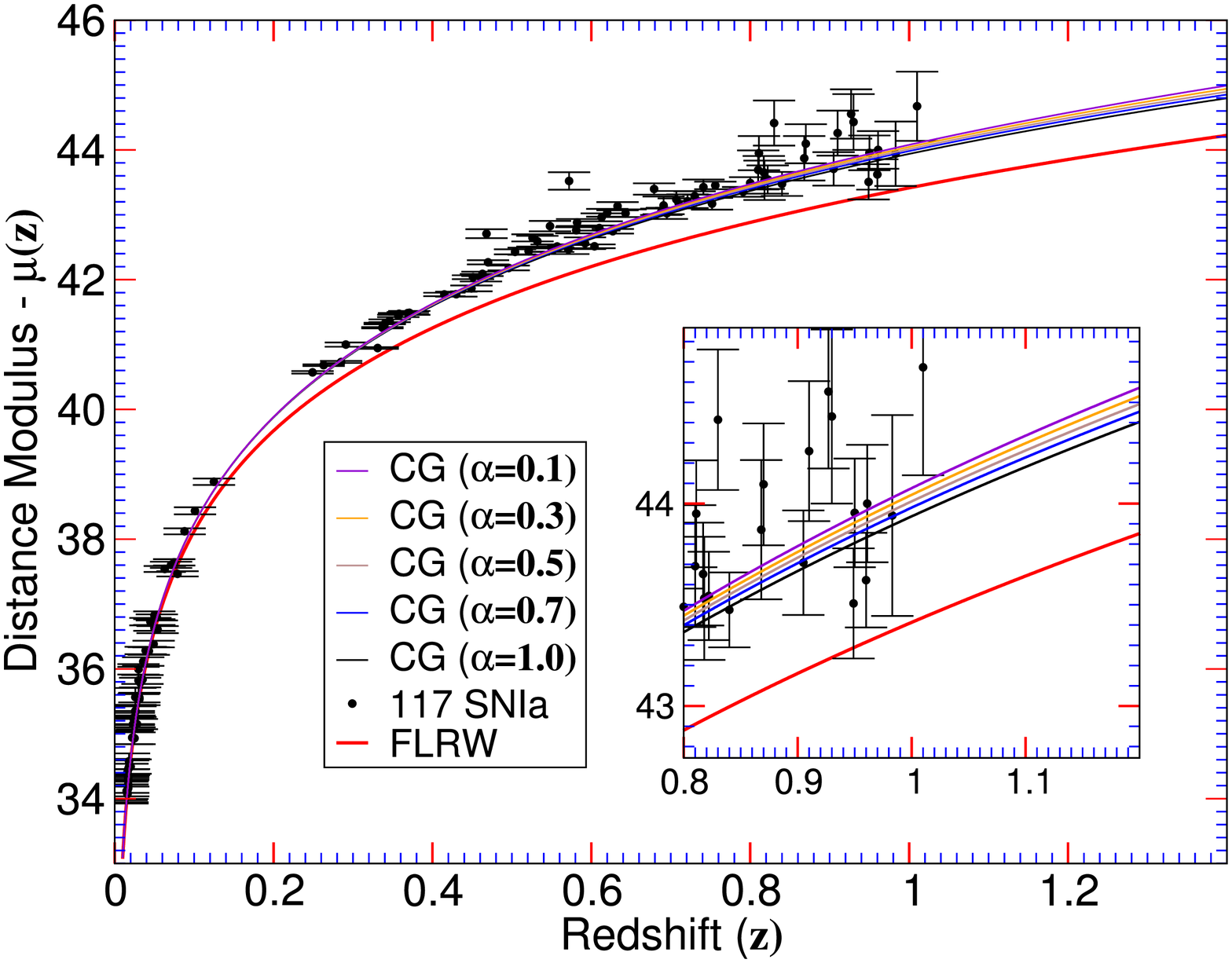}\hskip 0.50 truecm
\includegraphics[height=2.0in,width=2.0in]{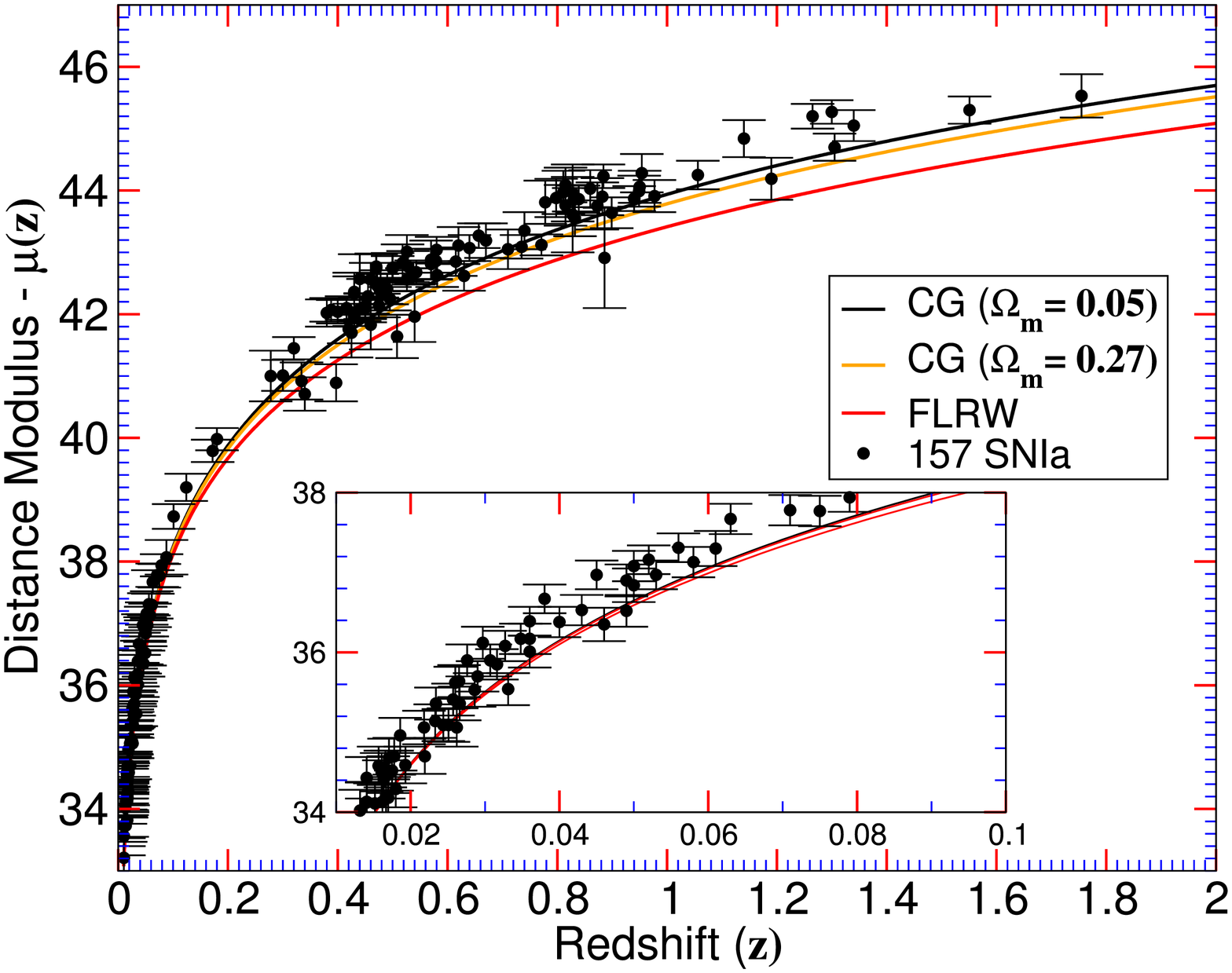} }%\hskip 0.30truecm 
\label{figure5}
\end{figure*}

\begin{figure}[tbh]
\centering{
%\resizebox{8.2cm}{!}{
%[angle=-90]
\includegraphics[height=2.0in,width=2.0in]{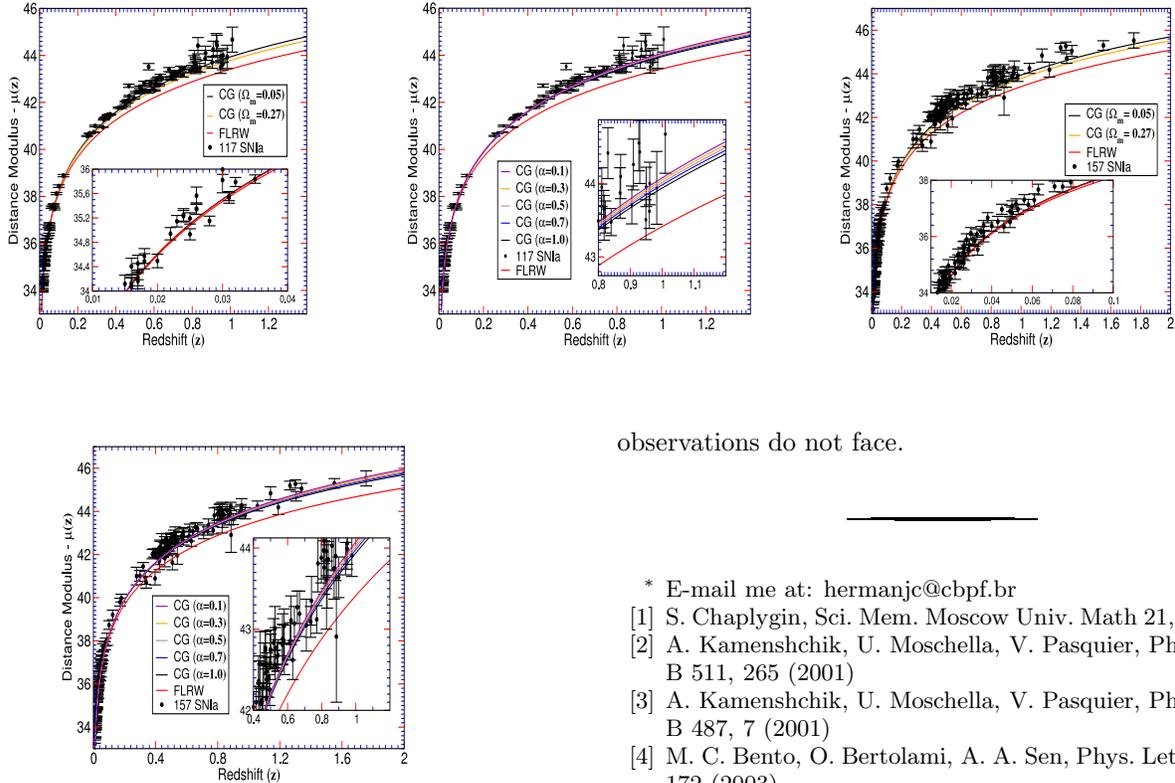} }
\caption{(color on-line) Hubble diagram of SNIa: GOLD sample (leftish pair) and LEGACY 
sample (rightish pair), and the prediction of the CG model of DE, including the case of a 
cosmological constant with $\Omega_\Lambda = 0.73$. Similar plots are also given for $\alpha 
= 0.1, 0.3, 0.5, 0.7$ .} 
\label{figure5}
\end{figure}

{\sl Data analysis and results.---} 
For the present analysis we have taken benefit of two samples of GRBs as compiled by 
Schaefer \cite{Schaefer2006}, 52 GRBs that were analyzed by five different methods, 
and by Bloom, Frail and Kulkarni \cite{Bloom-Frail-Kulkarni2003}, 24 GRBs. Both having 
properly estimated distance modulus, $\mu(z)$, and redshifts. We also used the SNIa data 
as collected in the GOLD sample \cite{riess2004}, and the SuperNova Legacy Survey (SNLS) 
\cite{astier2005}. Our main results are collected in Figures 1, 2 and 3 (color on-line). 
They were obtained upon integration of the luminosity distance function in 
Eqs.(\ref{luminosity-distance},\ref{luminosity-distance-function}), for a flat universe 
with $ \Omega_{m0} = 0.05$ and $\Omega_{c0} = 0.95$, and for a cosmological 
constant-like CG with $\Omega_{c0} = 0.73$ and $ \Omega_{m0} = 0.27$.

%\vskip 1.0 truecm

{\sl Conclusions.---} 
We presented the Hubble diagram of a large sample of GRBs having properly estimated 
redshifts together with the equivalent plot predicted by the cosmology of the Chaplygin 
gas, one of the various candidates to make-up the dark energy; the exotic fluid driving 
the universe late-time accelerated expansion, and FLRW and $\Lambda$-CDM models. After 
doing the cosmology of this model, we performed the statistical analysis and computed 
the model-to-sample $\chi^2$. It was found that the CG definitely fits much better the 
HD of GRBs used in the present study, in comparison to the Friedmann and the pure 
$\Lambda$-CDM models. The present GRBs HD plot exhibits a marked trend indicating that 
as one goes back in time, i.e., to very high redshifts, it gets much closer to the predict 
HD for a Friedmann universe. This clear convergence demonstrates that a transition from
decelerate to accelerate expansion did take place. However, and contrarily to claims based 
on supernovae type Ia (SNIa), the transition redshift lies somewhere between $2.5 \lesssim 
z \simeq 3.5$ rather than at $z \sim 0.5-1$. Therefore, this 
GRBs HD constitutes the most robust demonstration that the Chaplygin gas can indeed be the 
universe's driving dark energy field. ``En passin", as still the GOLD and LEGACY SNIa HD
locate far-above the CG predicted expansion law, this is further evidence of their 
intrinsic problems: or the {\sl Phillips relation} should be revisited 
\cite{middleditch06}, or perhaps there is some interrelation between the estimated 
luminosity distance residuals and internal extinction of the host galaxy, as some 
researchers have 
pointed out recently \cite{balazs2006}. Finally, it is worth to quote that a similar 
conclusion regarding the transition redshift $z \sim 3$ was achieved quite recently by 
Amendola, Gasperini and Piazza\cite{amendola06} upon the admission that dark matter and 
dark energy can strongly interact and evolve through a scaling regime $\rho_{\rm{DM}} 
\sim \rho_{\rm{DE}} \sim a^{-3(1 + \omega_{\rm{eff}})}$, with 
$\omega_{\rm{eff}}$ a constant. However, although the present analysis 
here do confirm the transition around $z \sim 3$, as we pointed out here, 
and in Refs.\cite{nos2006a,nos2006}, the GOLD and LEGACY SNIa observations 
seem to violate the general relativistic energy conditions (see Figure 3) 
for the (generalized) Chaplygin gas and Friedmann models, a problem that 
GRBs observations do not face.

\end{document}